# The Infrared Diameter – Velocity Dispersion Relation for Elliptical Galaxies

Michael D. Gregg

*Institute for Geophysics and Planetary Physics*
*Lawrence Livermore National Laboratory*
*L-413, 7000 East Avenue*
*Livermore, CA 94550*
*gregg@igpp.llnl.gov*

## ABSTRACT

Using single channel infrared photometry from the literature, a provisional K-band diameter–velocity dispersion relation for elliptical galaxies in the Coma and Virgo clusters is derived. The Coma cluster relation has ∼1.5 times lower scatter at K than in B or V. Excluding 4 outliers, the RMS scatter at K for 24 galaxies in Coma is only 4.8% in distance, close to the limit implied by the observational errors. Distance estimates based on the IR $D - \sigma$ relation will be more accurate than those derived from optical data. The improvement in the infrared is attributed to a decrease in sensitivity to stellar population parameters (age, metallicity, and slope of the IMF) as well as lower internal extinction from dust compared to the optical. That the $D - \sigma$ relation has a larger scatter in the optical indicates that there are detectable, but small, stellar population or dust content differences among the Coma ellipticals. Since the $D - \sigma$ relations are based on the fundamental plane, this result promises that the fundamental plane is thinner in the infrared than it is in the optical.

Infrared photometric data available for Virgo are limited to just 13 objects; the spread in distance due to the depth of the cluster precludes any significant improvement over B and V. A relative Coma-Virgo distance of 5.56 is derived from the K band data, in agreement with estimates in other colors and using other techniques, indicating that there is no significant age difference between Virgo and Coma ellipticals.

*Subject headings:* galaxies: distances — galaxies: elliptical

## 1. Introduction

There are two popular methods for determining redshift-independent distances to galaxies. Both rely on a tight correlation between a spectroscopically determined distance-independent measure of the gravitational potential and a photometrically determined distance-dependent measure of apparent size. The Tully-Fisher (TF) relation for spirals (Tully & Fisher 1977) is based on the correlation between luminosity and rotational velocity. The $D_n - \sigma$ technique for ellipticals (Dressler *et al.* 1987; D87) depends on the correlation between a surface brightness dependent diameter and the central velocity dispersion ($\sigma$).

When first introduced, the TF method employed B band photometry, necessitating relatively large and uncertain corrections for internal extinction which vary with galaxy morphology and star formation rate (Aaronson, Huchra, & Mould 1979 AHM). Differing current SFR's probably also contribute to the spread in B magnitudes. The accuracy of the TF method was improved markedly when AHM introduced the use of H band ($1.6\mu$) photometry where extinction corrections are reduced to negligible levels and the flux is mainly from the oldest population of stars which is perhaps more uniform from galaxy to galaxy (AHM).

There is increasing evidence from spectroscopic studies suggesting that the stellar population of ordinary elliptical galaxies can be a mix of ages (*e.g.*, O'Connell 1980; Rose 1985; Pickles 1985, Bower *et al.* 1990, Gonzales 1993). Extreme cases such as merger remnants have experienced strong bursts of star formation (Carter *et al.* 1988; Schweizer *et al.* 1990). Dust has also been shown to exist in a large fraction of ellipticals. There is good evidence that the distances derived from $D_n - \sigma$ and the closely related fundamental plane (FP) technique (Djorgovski & Davis 1987) for some ellipticals are influenced by one or more of these factors (Gregg 1992, 1995; Guzman & Lucey 1993). By analogy with the TF relation, it is natural to suspect that replacing B-band with infrared photometry might lead to an improvement in the accuracy of distances estimated by $D_n - \sigma$ and FP plane techniques (Guzman & Lucey 1993; Gregg 1995). This paper presents evidence that the $D - \sigma$ method, and by inference the FP, has significantly less scatter in the IR than in the optical for ellipticals in the Coma cluster.

## 2. Derivation of Infrared Diameters

Using IR data from Bower, Lucey, & Ellis (1992a; BLE), it has been shown elsewhere that the K-band ($2.2\mu$) magnitude-velocity dispersion relation for 25 early type galaxies in the Coma cluster has a scatter smaller by a factor of 1.9 than the B band $D - \sigma$ relation (Gregg 1995). Because the IR magnitudes are from single channel photometry in a 17" aperture, the derived IR relation is between velocity dispersion and a *metric* size. For a useful distance indicator, it is necessary to define a size based on structural parameters of the galaxies individually; the FP or $D - \sigma$ relations accomplish this using the half light-radius $R_e$ or $D_n$. D87 define $D_n$ as the diameter of a circular aperture within which the mean B band surface brightness is 20.75 mag/sq arcsec and show that $\log(D_n)$ is well correlated with $\log(\sigma)$, reducing the scatter by a factor of two from the Faber-Jackson ($B_T$ vs. $\log(\sigma)$) relation (Faber & Jackson 1976). Typical ellipticals have B-K $\sim 4.0$, so a K band diameter ($D_K$) which approximates $D_n$ would enclose a mean K surface brightness of 16.75 mag/sq arcsec.

The Virgo and Coma $D_n$ are based on CCD photometry (D87; Burstein *et al.* 1988, B88). The ideal IR version of a $D - \sigma$ relation would use K band imaging [1]; however, with some simple as-

---

[1] Jorgensen, Franx, & Kjaergaard (1993) have demonstrated that the FP is an intrinsically more accurate distance indicator than $D - \sigma$. Using the fits detailed here, it is possible to derive an IR version of the FP. Because

sumptions a provisional $D_K - \sigma$ relation can be derived for the Coma and Virgo clusters from the data of Persson, Frogel, & Aaronson (PFA 1979) and BLE; both present high quality single channel IR photometry for early type galaxies in the Coma and Virgo clusters. The BLE photometry has been calibrated to be on the same system as that of PFA (see BLE for discussion) and their much larger body of photometry supersedes the earlier PFA work. As BLE present no new IR photometry for Virgo, the data of PFA are used here.

Table 5 of BLE lists K magnitudes through an effective aperture of 17" for 25 E or E/S0 and 3 S0 galaxies in the Coma cluster which also have $\log(\sigma) > 2.0$ in Davies *et al.* (1988) or Dressler (1987). Corrections for background light, redshift, and Galactic reddening were applied as detailed in BLE and PFA. $D_K$ can be derived from these single aperture data by assuming an $R^{1/4}$ law profile and adopting an effective diameter $A_e$ from the V band photometry of Lucey *et al.* (1991a LGCT) for the ellipticals and the *Third Reference Catalog* (de Vaucouleurs *et al.* 1991) for the S0's. In the absence of color gradients, $A_e$ is independent of wavelength and in most ellipticals, color gradients are negligible (PFA; Peletier, Valentjin, & Jameson 1990).

The $R^{1/4}$ fits are uniquely determined by the aperture photometry and effective diameter. Errors have been estimated by deriving $D_K$ after varying both quantities by their estimated standard deviations, $\pm 0.028$ dex for $log(A_e)$ (LGCT) and $\pm 0.026$ mag for the K aperture magnitudes (BLE). These small errors have little effect on $\log(D_K)$, amounting to only $\pm 0.01$ dex RMS,

---

of the use single channel IR data and $A_e$ from V band measurements, this is not preferred here because the V band photometry then plays a role in determining 2 of the 3 dimensions of the FP ($A_e$ and $\mu_e$), and contributes to both sides the FP distance estimator itself, $A_e$ as well as $\log(\sigma)+C*\mu_e$. The results, however, of such an analysis are in agreement with the $D-\sigma$ relation discussed here. Several groups are now working towards a full K-band FP derivation (*e.g.*, Pahre *et al.* 1995).

smaller than the error estimates for $\log(D_n)$ Burstein *et al.* (1987) and comparable to those in $\log(D_V)$ (LGCT). The derived $\log(D_K)$ are listed in Table 1 along with their $A_e$, $D_n$ from B88, and the V band equivalent, $D_V$ (aperture diameter enclosing 19.80 mag/sq arcsec) from BLE. Note that although V band $A_e$ from LGCT have been used in deriving $D_K$, strict photometric independence between the V and K band results is maintained by adopting $D_V$ from BLE.

PFA present JHK photometry of 13 Virgo ellipticals through 2 apertures (29" and 56"), which allows $D_K$ to be derived from an $R^{1/4}$ law fit without recourse to any optical data. For comparison, $D_K$ values were also derived for the Virgo ellipticals using the same method as the Coma ellipticals, *i.e.*, by adopting an optical $A_e$ and fitting to a single aperture. It is testimony to the high quality of the data from PFA, BLE, and B88 that the two methods give identical results within a few percent and that these are in turn extremely well correlated with the diameters in B and V. This test serves as a confidence check on the derived IR diameters for the Coma galaxies. Table 1 includes the 13 Virgo ellipticals from PFA and their calculated $D_K$, along with $D_n$ (B88) and $D_V$ (BLE). One of the objects, NGC 4387, with $\log(\sigma) < 2.0$, is excluded from the analysis to follow.

### 2.1. Velocity Dispersions

Velocity dispersions from Davies *et al.* (1987) have been adopted because these data produce tighter $D - \sigma$ correlations than other data sets (*e.g.*, Dressler 1984 or LGCT). This can be attributed to the careful treatment of systematic effects as well as the averaging of several independent measurements in many cases. In 3 cases, however, the velocity dispersions of LGCT are preferable. The LGCT $\sigma$ for RB 257 is more than 4 rms lower than that of Davies *et al.* and moves RB 257 directly onto the $D - \sigma$ relation in all colors. The Davies *et al.* dispersion for NGC 4874 is only 245 km/s, well below other es-timates (*c.f.* Whitmore, McElroy, & Tonry at 304 km/s). The intermediate $\log(\sigma)$ of LGCT (279 km s$^{-1}$) again moves NGC 4874 onto the $D-\sigma$ relation in all bands. Finally, the Davies *et al.* value for NGC 4923 is the only one in the present sample based only on lower resolution Lick data; the higher precision LGCT value is superior. By adopting the alternative velocity dispersions for these three objects, the scatter in the $D-\sigma$ relation is minimized in all three colors.

Table 1
Data for Coma and Virgo Galaxies

| Name | D# | $\sigma$ | $A_e$ | $D_n$ | $D_V$ | $D_K$ |
|---|---|---|---|---|---|---|
| \multicolumn{7}{c}{Coma Cluster} | | | | | | |
| NGC 4839 | 31 | 2.413 | 1.725 | 1.238 | 1.278 | 1.306 |
| NGC 4926 | 49 | 2.431 | 1.281 | 1.248 | 1.269 | 1.293 |
| IC 3959 | 69 | 2.301 | 1.060 | 1.048 | 1.064 | 1.093 |
| IC 3957 | 70 | 2.176 | 0.921 | 0.938 | 0.948 | 0.932 |
| NGC 4923 | 78 | 2.318 | 1.147 | 1.128 | 1.144 | 1.137 |
| NGC 4906 | 118 | 2.225 | 1.149 | 1.058 | 1.029 | 1.030 |
| NGC 4876 | 124 | 2.260 | 0.990 | 1.008 | 1.030 | 1.028 |
| RB 43 | 125 | 2.221 | 0.515 | 0.860 | 0.874 | 0.825 |
| NGC 4874 | 129 | 2.446 | 2.040 | 1.288 | 1.287 | 1.324 |
| NGC 4872 | 130 | 2.326 | 0.768 | 1.058 | 1.048 | 1.054 |
| NGC 4867 | 133 | 2.346 | 0.852 | 1.058 | 1.060 | 1.070 |
| RB 257 | 136 | 2.130 | 0.446 | 0.868 | 0.877 | 0.865 |
| IC 4051 | 143 | 2.348 | 1.572 | 1.078 | 1.107 | 1.134 |
| NGC 4889 | 148 | 2.581 | 1.775 | 1.468 | 1.481 | 1.544 |
| IC 4011 | 150 | 2.025 | 0.976 | 0.868 | 0.868 | 0.854 |
| NGC 4886 | 151 | 2.215 | 1.173 | 1.048 | 1.036 | 1.006 |
| IC 3998 | 152 | 2.192 | 1.18 | 0.895 | 0.903 | 0.934 |
| RB 45 | 153 | 2.130 | 0.813 | 0.858 | 0.842 | 0.863 |
| IC 4045 | 168 | 2.324 | 0.960 | 1.098 | 1.123 | 1.148 |
| IC 4026 | 170 | 2.146 | 1.16 | 0.917 | 0.921 | 0.874 |
| IC 4021 | 172 | 2.199 | 0.777 | 0.958 | 0.956 | 0.955 |
| IC 4012 | 174 | 2.253 | 0.577 | 0.978 | 0.999 | 1.032 |
| NGC 4883 | 175 | 2.234 | 1.00 | 1.026 | 1.027 | 1.050 |
| RB 155 | 193 | 2.079 | 0.858 | 0.818 | 0.828 | 0.808 |
| NGC 4860 | 194 | 2.391 | 1.185 | 1.168 | 1.192 | 1.228 |
| RB 167 | 207 | 2.167 | 0.885 | 0.898 | 0.894 | 0.894 |
| NGC 4881 | 217 | 2.340 | 1.293 | 1.108 | 1.134 | 1.146 |
| NGC 4841A | 240 | 2.422 | 1.459 | 1.248 | 1.299 | 1.302 |
| \multicolumn{7}{c}{Virgo Cluster} | | | | | | |
| NGC 4365 | ⋯ | 2.394 | ⋯ | 1.88 | 1.88 | 1.96 |
| NGC 4374 | ⋯ | 2.458 | ⋯ | 2.03 | 2.01 | 2.07 |
| NGC 4387 | ⋯ | 1.924 | ⋯ | 1.48 | 1.47 | 1.47 |
| NGC 4406 | ⋯ | 2.398 | ⋯ | 2.02 | 1.94 | 2.02 |
| NGC 4458 | ⋯ | 2.025 | ⋯ | 1.39 | 1.39 | 1.37 |
| NGC 4472 | ⋯ | 2.458 | ⋯ | 2.14 | 2.14 | 2.23 |
| NGC 4473 | ⋯ | 2.250 | ⋯ | 1.87 | 1.86 | 1.88 |
| NGC 4478 | ⋯ | 2.174 | ⋯ | 1.70 | 1.67 | 1.69 |
| NGC 4486 | ⋯ | 2.558 | ⋯ | 2.08 | 2.09 | 2.21 |
| NGC 4552 | ⋯ | 2.417 | ⋯ | 1.93 | 1.90 | 1.97 |
| NGC 4621 | ⋯ | 2.381 | ⋯ | 1.91 | 1.90 | 1.94 |
| NGC 4636 | ⋯ | 2.281 | ⋯ | 1.86 | 1.87 | 1.92 |
| NGC 4660 | ⋯ | 2.296 | ⋯ | 1.69 | 1.71 | 1.73 |

Additional support for these three velocity dispersions being superior comes from the fundamental plane: in planar fits to $\log(\sigma)$ and $\mu_e$ of the other galaxies in B and V, the alternative dispersions lower the residuals of the three objects in question, with the exception of RB 257 in the B band, by 2-5$\sigma$. The effective radius of RB 257 varies by a factor of more than 1.5 in the literature, ranging from 1.4" to 2.2". Seeing effects may be dominating the errors in R$_e$ for this object and its effective radius must be considered uncertain.

## 3. Coma

The IR and optical $D-\sigma$ relations for the Coma objects are compared in Figure 1 (upper panels, a-c). The filled circles are ellipticals and the 4-pointed stars are S0's. The open circles are 3 of the highest surface brightness galaxies in the sample (RB 43, NGC 4867, and NGC 4872). These clearly lie off the trend in K and tend to be high in V and B as well. They will be discussed further below and are not used in the fits. The "+" is IC 4011 which has a $\log(\sigma)$ just above the cutoff of 2.0 applied by D87; it lies well below the relation in all 3 colors and so will be excluded from further consideration. The lines through the data in Figure 1 are derived from a robust fitting technique based on median values (Emerson & Hoaglin 1983), with $\log(\sigma)$ as the independent parameter, to the 24 remaining galaxies. Due to the nature of the robust fitting technique, there is little difference between fits with and without the outliers. The S0 galaxies fit well among the ellipticals, as demonstrated by Dressler (1987). The best fits and Spearman correlation coefficients and probabilities are

$$\text{Log}(D_n) = 1.279 \; \log(\sigma) - 1.855; \quad (1)$$
$$S = 0.96; P_S = 7.1 \times 10^{-14}$$

$$\text{Log}(D_V) = 1.319 \; \log(\sigma) - 1.938; \quad (2)$$
$$S = 0.97; P_S = 4.93 \times 10^{-15}$$

and

$$\text{Log}(D_K) = 1.458 \; \log(\sigma) - 2.243; \quad (3)$$
$$S = 0.98; P_S = 9.26 \times 10^{-18}.$$

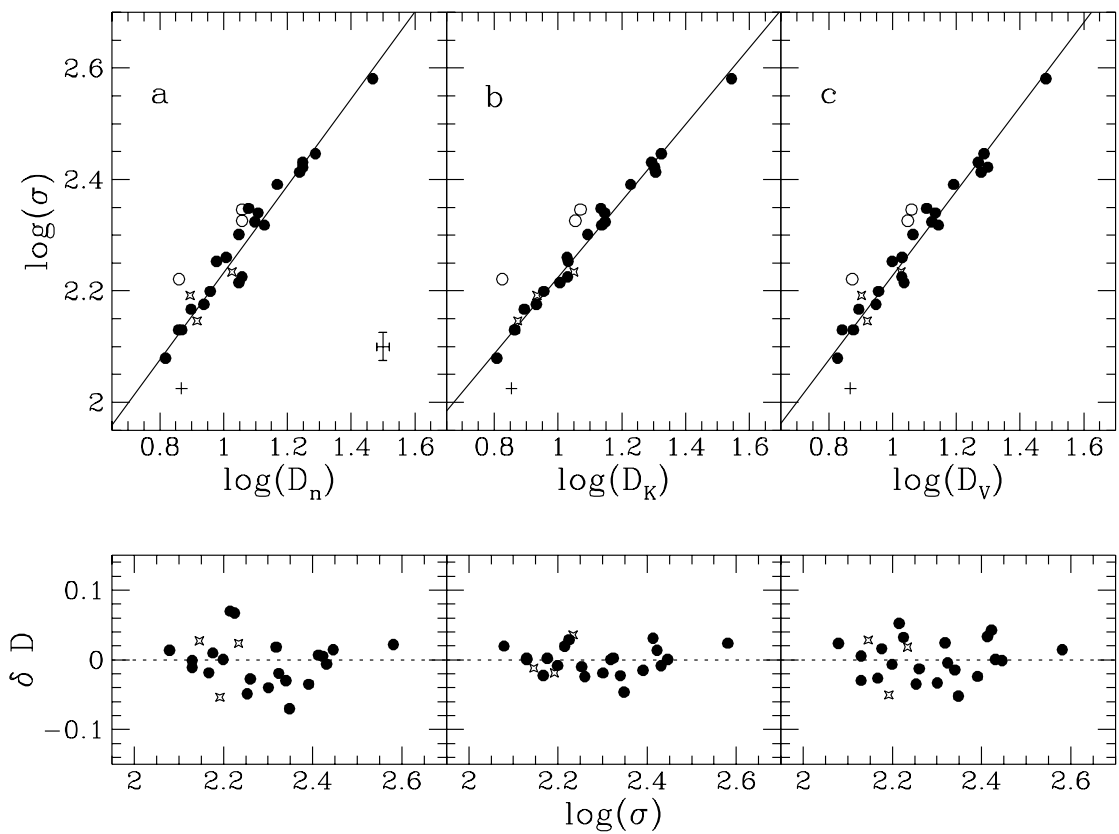

Fig. 1.— Upper panels: comparison of the B, V, and K $D - \sigma$ relations for the Coma cluster. Filled circles are ellipticals, 4-pointed stars are S0's, open circles are outliers identified in the K band data (see text). The "+" is IC 4011. Lines are robust fits to the filled circles and 3 S0's. Lower panels: residuals of the 24 E and S0 objects used in the fitting. The RMS in K is 1.68 times lower than in B, 1.42 times lower than in V. A representative error bar is plotted showing ±0.02 in Log(D). The error estimates for Log($D_K$) and Log($D_V$) are half this amount.

The residuals in Log(D) at a given velocity dispersion are plotted in the lower panels of Figure 1.

Immediately apparent from Figure 1 is that the $D_K - \sigma$ relation is tighter than either of its optical counterparts. The RMS scatter of the 24 galaxies is 0.0345, 0.029, and 0.0205 in log(D) for B, V, and K, respectively. The K band relation reduces the scatter by a factor of 1.68 from B and 1.42 from V. This corresponds to a scatter in distance of only 4.8%, roughly the predicted limit from measurement errors (Jorgensen et al. 1993). The RMS for B and V are both less than that found by Jorgensen et al. from the Coma

FP for a similar sample; this can be attributed to the use of the alternative velocity dispersions and removal of the 3 remaining outliers. It can be demonstrated that removing the outliers and using the better velocity dispersions also improves the FP fits.

In K, there are three objects which lie well above the tight relation defined by the majority. At V, these objects are less clearly outliers and at B, only RB 43 is clearly away from the rest, due to the increase in sample RMS. The trend with bandpass suggests stellar population effects, but these three objects have typical V-K (BLE). These 3 are among the 5 highest surface brightness objects in the sample. Jorgensen *et al.* have shown that the correlation of residuals in the $D-\sigma$ relation with surface brightness (Lucey, Bower, & Ellis 1991) results from the $D-\sigma$ relation being a not-quite-optimum projection of the FP. While this may account for these outliers, it is noted that RB 257 and IC 4012 also have surface brightnesses in the same range, yet lie close to the $D-\sigma$ relations in all colors (though not if $\log(\sigma)$ from Davies *et al.* is adopted for RB 257). LGCT redetermined $\log(\sigma)$ for RB 43 and it is in good agreement with the Davies *et al.* value. Unusual central dynamics may be responsible, but further longslit observations are needed to resolve this (Jorgensen *et al.* 1993). An independent, comprehensive spectroscopic study at high signal-to-noise and high dispersion is needed to place the Coma velocity dispersions on a truly uniform system.

### 3.1. Statistical Significance

The F-test indicates that the sample variance of $D_K$ is significantly smaller than $D_n$ at the 98.5% confidence level, while the improvement over $D_V$ is significant at the 91.0% level. The precision of the photometry is such that the velocity dispersions are the limiting factor and contribute identically to the scatter in all three colors, resulting in highly correlated residuals. The F-test therefore underestimates the real significance of the differences. A Student's 2 point t-test takes this into account (Press *et al.* 1984) and indicates that the average absolute value of the residuals in $D_K$ is significantly lower than in $D_n$ or $D_V$ at the 99.2% and 98.9% confidence levels, respectively. Either test applied between $D_V$ and $D_n$ shows that they are not significantly different at the ~45% confidence level. The conclusion is that $D_K - \sigma$ is the best relation at a significantly higher level.

### 3.2. Why is K better?

Jorgensen *et al.* (1993) show that the observed scatter in the Coma FP in B and r is about twice the expected from the measurement errors, but could not identify a significant additional parameter, ruling out ellipticity, B-r colors or gradients, isophotal twists, boxiness, and the presence of a disk. The photometric errors in B, V, and K are 0.02-0.03 magnitudes; since the velocity dispersions in Figure 1 are identical in B, V, and K, the increase in width of the $D-\sigma$ relation with decreasing wavelength is due to "cosmic" effects. One possible source of this additional scatter, which explains the decrease with wavelength, is variations of stellar population parameters (age, IMF slope, or metallicity) or the amount of dust (which may be linked to the stellar population) at a given velocity dispersion. All of these will have a larger effect in the optical than in the IR. Modeling carried out by Tonry, Ajhar, & Luppino (1990) shows that the I band is less affected by these variables than V or R, and it is reasonable to expect that the sensitivity is further reduced at K. Models by Bruzual (1983) suggest that this is the case (Guzman & Lucey 1993). The real situation may be a good deal more complicated, however. Early results from the surface brightness fluctuation distance indicator technique applied in the K band show a *larger* scatter galaxy-to-galaxy than in the I band (Luppino & Tonry 1993; Pahre & Mould 1994). If further observations of nearby ellipticals continue to exhibit a large range in IR properties, the improvement in the $D-\sigma$ relation demonstrated here for Coma in the K band may not be universal.

The variations in age, IMF slope, and metallicity can be quite small and still account for the observed increase in scatter. For instance, the total scatter in the B band $D-\sigma$ relation can be accounted for by a corresponding RMS in age of only 0.6-0.9 Gyr for simple models with mean age of 15 Gyr, typical IMF, and solar or twice solar metallicity (Gregg 1995, Table 1). Quite small changes in the metallicity or IMF can just as readily account for the scatter. This is an upper limit since some portion of the scatter is due to observational error rather than cosmic differences; taking the observational errors into account lowers the age scatter by 20-30%. The presence of dust would further reduce the allowable scatter in the other stellar population parameters. From these data alone, however, it is impossible to determine whether metallicity, age, the initial mass function, or simply dust is responsible for the additional scatter in the optical.

Similarly tight constraints on the ages have been reached by Bower, Lucey, & Ellis (1992b) using these same data in the form of color-magnitude relations. Although they detect a scatter in their Coma V-K color-magnitude relations larger than their estimated errors, they are reluctant to attribute it to a real variation in the Coma galaxies. The differences seen in the $D-\sigma$ relations of Figure 1 strongly argue that the larger scatter in the optical is due to intrinsic variations among the Coma early type galaxies at the tiny levels discussed above.

Such tight constraints on the spread in the stellar population parameters stands in stark contrast to the often large range in ages attributed to ellipticals in the spectroscopic studies cited in the Introduction. The disagreement can be resolved, as already suggested above, if the Coma cluster ellipticals contain a highly uniform stellar population. Because most of the spectroscopic population studies have been conducted with narrow slits on relatively nearby objects, aperture effects may also play a role. If the younger populations are confined to the inner 1-2" in galaxies 5-20 Mpc distant, they may be undetectable in Coma galaxies provided they have the same physical extent. A comparison of $H\beta$ line strengths in the Coma and Virgo clusters (based on the same data as was used in the velocity dispersion determinations) by Gregg & Dressler (1986) shows that differences do exist and is consistent with an aperture effect. Bower *et al.* (1990), however, argue against aperture effects as an explanation for the spectroscopic differences they find in a comparison of ellipticals from the Virgo and Coma clusters and other environments. A much more detailed spectroscopic comparison of ellipticals in the Virgo and Coma cluster is called for to shed more light on this issue.

Since the scatter in the $D-\sigma$ relation is smaller at K, then at a given luminosity or velocity dispersion, the residuals in the optical $D-\sigma$ relation should correlate with galaxy color. Galaxies which have too small a diameter in the optical at a given velocity dispersion must be brighter in the IR and hence redder for their total luminosity; galaxies which are too large in the optical will be bluer. Thus the diameter residuals from the $D-\sigma$ relation should *anticorrelate* with the residuals from the color-magnitude relation. Figure 2 plots the 24 residuals $\delta D_n$ of Figure 1 against the residuals from the $V_T$—(V-K) relation of Bower *et al.* (1992b), based on a sample of 40 E+S0 Coma galaxies. The Spearman rank correlation coefficient for the residuals is -0.63, probability $1.0 \times 10^{-3}$. The residuals in this plot are completely independent of each other and the evident anticorrelation strongly supports the results of Figure 1 that the $D-\sigma$ relation is tighter in the IR and is consistent with the interpretation that the additional scatter in the optical is from stellar population variations or dust.

Jorgensen *et al.* (1993) report no correlation of residuals from a FP fit with B-r color resid-uals at a given velocity dispersion for the Coma galaxies. The positive result shown in Figure 2 can be attributed primarily to greater sensitivity provided by the longer wavelength baseline of V-K, as well as complete independence of the two residual quantities. Correlated errors in residuals both involving $\log(\sigma)$ will tend to wash out any real effects from other causes.

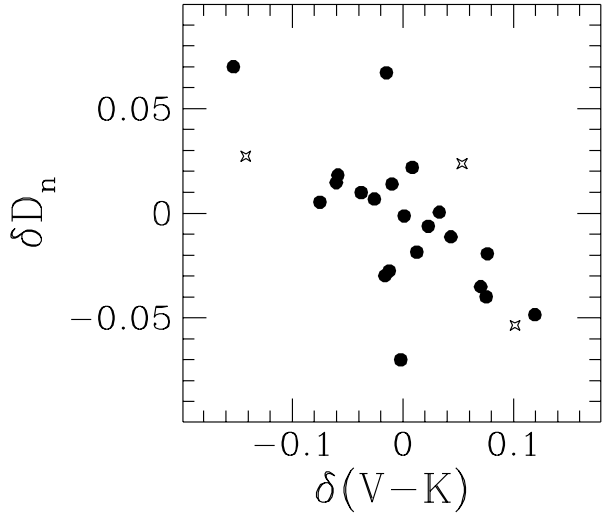

Fig. 2.— Residuals $\delta D_n$ from the B band relation of Figure 1 plotted against residuals $\delta(V-K)$ from the V band color - magnitude relation of Bower et al. (1992b). The anti-correlation is expected if either stellar populations or dust are responsible for the increase in scatter of the B band $D_n - \sigma$ relation. Symbols as in Figure 1.

Even if the use of optical effective diameters to derive $D_K$ for the Coma sample is inappropriate and the results of Figure 1 are completely fortuitous, Figure 2 is an independent prediction that IR imaging data will show that the fundamental plane is thinner at K than in the optical. An IR imaging study of ellipticals in Coma, Virgo, and other clusters is needed for an independent assessment of the total improvement available at K and a complete characterization of the FP in the IR. Such work is already being carried out by several groups (e.g., Pahre et al. 1995).

## 4. Virgo

Figure 3 compares the $D_n - \sigma$ and $D_V - \sigma$ relations for 13 Virgo ellipticals with the $D_K - \sigma$ relation derived here, based on the PFA data. The lines are linear least squares fits, forcing the same slope as derived from the Coma relations, excluding NGC 4387 ("+" sign) because of its low velocity dispersion. There is little difference among the colors and the RMS residuals are 2-3 times those for the Coma cluster. Within the uncertainties due to the large scatter and small number of objects, the Virgo relations are compatible with those for Coma. The larger scatter of the Virgo ellipticals and the lack of improvement going from B to K can be attributed primarily to the sizable depth of the Virgo cluster (e.g. Tonry et al. 1990). The computed intercepts for B, V, and K are -1.121, -1.224, and -1.498, respectively. From these, the relative Coma-Virgo distance in each color can be derived: 5.42 in B, 5.18 in V, and 5.56 in K, essentially identical given the scatter of $\sim 0.07$ in Log(D). These relative distances are in good agreement with other estimates (D87; Sandage & Tammann 1990; Bower et al. 1992b).

Bower et al. (1990) present evidence for a rather large age difference between the elliptical populations of the 2 clusters, based on Rose's (1985) spectral indices applied to composite spectra. The indicated age difference is approximately a factor of two which in turn predicts that the derived distance ratio should be a rather strong function of wavelength, being *smaller at longer wavelengths* (Gregg 1995). No such trend is found, arguing for *no significant difference in the mean ages of the Coma and Virgo ellipticals.* A similar conclusion regarding the relative ages has been reached by Bower et al. (1992b) based on broad band colors of ellipticals in the 2 clusters.

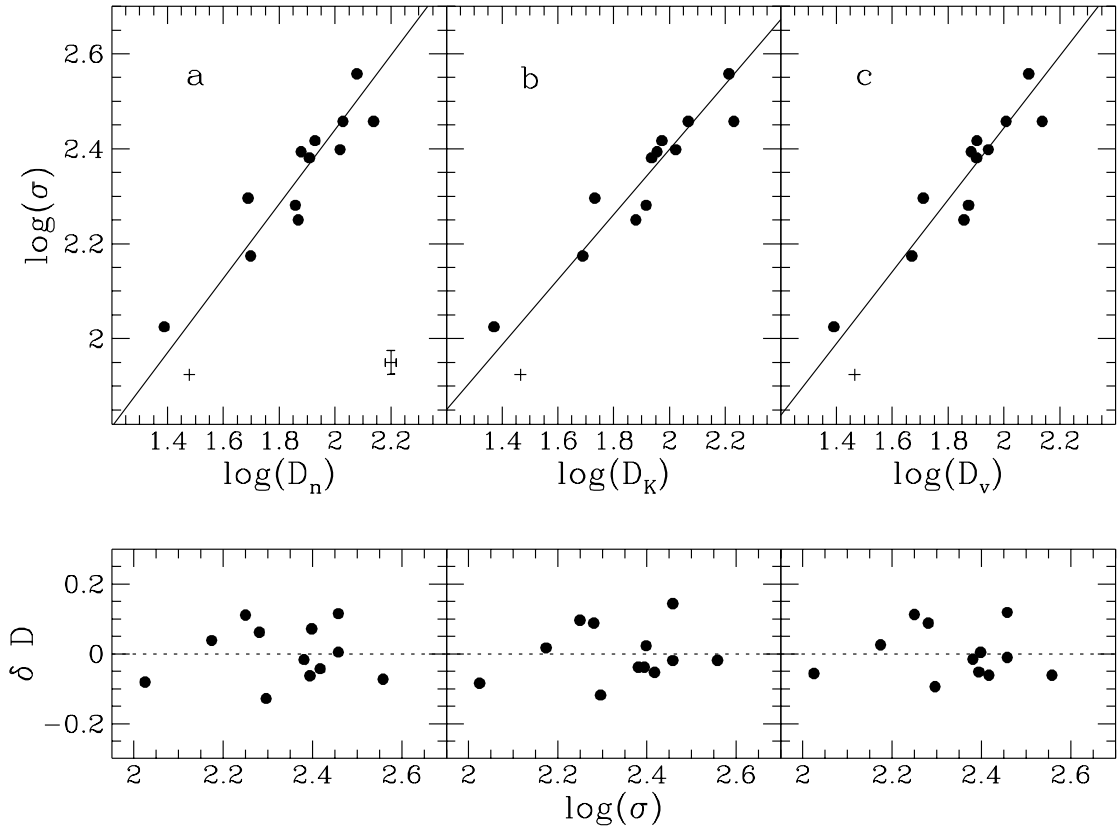

Fig. 3.— Upper panels: comparison of the B, V, and K D − σ relations for the Virgo cluster. The "+" is the low σ object NGC 4387. The lines are linear least squares fits (excluding NGC 4387), forcing the same slopes as in Figure 1 for the Coma cluster. Lower panels: residuals of the 12 E galaxies. The large scatter and high degree of similarity is attributed to the depth of the Virgo cluster. A representative error bar is plotted showing ±0.02 in Log(D). The error estimates for Log($D_K$) and Log($D_V$) are half this amount.

## 5. Summary and Conclusions

Using published single channel K band photometry combined with optical effective diameters and the assumption of an $R^{1/4}$ law, a provisional IR version of the diameter—velocity dispersion relation has been derived for early type galaxies in the Coma and Virgo clusters. The K band relation has 1.68 (1.42) times less scatter than the B (V) band D − σ relation for 24 galaxies in the Coma cluster, excluding a few outlying galaxies. The RMS scatter in distance is only 4.8% per galaxy for the Coma cluster K band D−σ relation, close to the limit from current observational errors alone (Jorgensen *et al.* 1993).

The improvement is attributed to lower sensitivity at K to variations in stellar population parameters (age, metallicity, and slope of the IMF) as well as negligible influence from any dust. Distance estimates based on K band photometry will be of superior accuracy to those in the optical. The improvement in the IR also indicates that there are small but detectable differences in either the stellar populations or the amount of dust in the Coma ellipticals. Because the D − σ relation depends on the FP, a further implication is that the FP is thinner in the IR than in the optical.

A relative Coma-Virgo distance of 5.56 is derived from the K band relation, in good agreement with estimates from the B and V bands and other estimates from a variety of techniques. This argues against any significant systematic age difference between the elliptical galaxy populations of the two clusters.

An IR imaging study of the Coma and Virgo ellipticals is needed to confirm the results derived here and to further investigate the FP at longer wavelengths. Additional velocity dispersion work at high resolution and S/N is needed to reduce the uncertainties in the velocity dispersion measurements which are the limiting factor in $D - \sigma$ and FP relations at all wavelengths. IR imaging and spectroscopy of other rich clusters would help to determine if the improvement seen in the Coma cluster K band $D - \sigma$ relation occurs in other environments and comparison with other colors could sensitively check the universality of $D - \sigma$ and FP relations.

# REFERENCES

Aaronson, M., Huchra, J., & Mould, J. 1979, ApJ, 229, 1

Bower, R.G., Ellis, R.S., Rose, J.A., Sharples, R.M. 1990, AJ, 99, 530

Bower, R.G., Lucey, J.R., & Ellis, R.S. 1992a, MNRAS, 254, 589 (BLE)

Bower, R.G., Lucey, J.R., & Ellis, R.S. 1992b, MNRAS, 254, 601

Bruzual, A.G. 1983, ApJ, 273, 105

Burstein, D., Davies, R.L., Dressler, A., Faber, S.M., Stone, R.S.P., Lynden-Bell, D., Terlevich, R.J. & Wegner, G. 1987, ApJS, 64, 601

Davies, R.L., Burstein, D., Dressler, A., Faber, S.M., Lynden-Bell, D., Terlevich, R.J. and Wegner, G. 1987, ApJS, 64, 581

Djorgovski, S. & Davis, M. 1987, ApJ 313, 65

Dressler, A. 1984, ApJ 281, 512

Dressler, A., Lynden-Bell, D., Burstein, D., Davies, R.L., Faber, S.M., Wegner, G., & Terlevich, R. 1987, ApJ, 313, 42

Dressler, A. 1987, ApJ 317, 1

Emerson, J.D., and Hoaglin, D.C. 1983, in Understanding Robust and Exploratory Data Analysis, ed. D.C. Hoaglin, F. Mosteller, and J.W. Tukey (New York: John Wiley & Sons, Inc.) p. 129

Faber, S.M. & Jackson, R.E. 1976, ApJ, 204, 668

Gonzales, J.J. 1993, Ph.D. Thesis, University of California, Santa Cruz

Gregg, M. & Dressler, A. 1986, in IAU Symp. 127, Structure and Dynamics of Elliptical Galaxies, ed. T. de Zeeuw (Dordrecht: Kluwer), 387

Gregg, M.D. 1992, ApJ 384, 43

Gregg, M.D. 1995, ApJ 443, 527

Guzman, R., & Lucey, J.R. 1993, MNRAS, 263, L47

Jorgensen, I., Franx, M., & Kjaergaard, P. 1993, ApJ, 411, 34

Lucey, J.R., Bower, R.G., Ellis, R.S. 1991b, MNRAS, 249, 755

Lucey, J.R., Guzman, R., Carter, D., & Terlevich, R.J. 1991a, MNRAS, 249, 755 (LGCT)

Luppino, G.A., & Tonry, J.L. 1993, ApJ, 410, 81

Lynden-Bell, D., Faber, S.M., Burstein, D., Davies, R.L., Dressler, A., Terlevich, R.J. & Wegner, G. 1988, ApJ, 326, 19

O'Connell, R.W. 1980, ApJ, 236, 430

Pahre, M.A., & Mould, J.R. 1994, ApJ, 433, 567

Pahre, M.A., Djorgovski, S.G., de Carvalho, R.R., & Mould, J.R. 1995, BAAS, 26, 1497

Peletier, R.F., Valentjin, E.A., & Jameson, R.F. 1990, AA, 233, 62

Persson, S.E., Frogel, J.A., Aaronson, M. 1979, ApJ 39, 61

Pickles, A.J. 1985, ApJ, 296, 340

Press, W.H., Flannery, B.P., Teukolsky, S.A., & Vettering, W.T. 1984, Numerical Recipes, The Art of Scientific Computing (Cambridge: Cambridge University Press)

Rose, J.A. 1985, AJ 90, 1927

Sandage, A. & Tammann G.A. 1990, ApJ 365, 1

Schweizer, F., Seitzer, P., Faber, S.M., Burstein, D., Dalle Ore, C.M., & Gonzales, J.J. 1990, ApJL, 364, 33

Tonry, J.L., Ajhar, E.A., & Luppino, G.A. 1990, AJ, 100, 1416

Tully, R.B., & Fisher, J.R. AA, 54, 661

Vaucouleurs, G. de, Vaucouleurs, A. de, Corwin, H.G., Buta, R.J., Fouqué, P., & Paturel, G. 1991 *Third Reference Catalog of Bright Galaxies*

Whitmore, B.C., Mcelroy D.B., Tonry J.L. 1985, ApJS, 59, 1

---